\documentclass[preprint,tightenlines,eqsecnum,floats,aps,amsmath,amssymb,nofootinbib,prd,showpacs]{revtex4}

\usepackage{amssymb}
\usepackage{stmaryrd}
\usepackage{amsmath}
\usepackage{amsfonts}
\usepackage{mathrsfs}
\usepackage{CJK}
\usepackage{float}
\usepackage{caption}
\usepackage{amsmath,amssymb,amsfonts}
\usepackage{graphicx}
\usepackage{subfigure}
\usepackage{color} 
\usepackage{fancyhdr}
\usepackage{hyperref} 
\begin{document}

\title{Gauss-Bonnet solution with a cloud of strings in de Sitter and anti-de Sitter space}
	
\author{Wenxi Zhai}
\affiliation{Department of Physics, South China University of Technology, Guangzhou 510641, China}
\author{ Xiangdong Zhang\footnote{Corresponding author. scxdzhang@scut.edu.cn}}
\affiliation{Department of Physics, South China University of Technology, Guangzhou 510641, China}
	
\date{\today}
	
\hyphenpenalty=5000
\tolerance=1000

\begin{abstract}
In this paper, we present exact spherically symmetric Gauss-Bonnet black hole solutions surrounded by a cloud of strings fluid with cosmological constant in $D>4$ dimensions. Both charged and uncharged cases are considered. We focus on the de Sitter solutions in the main text and leave the Anti-de Sitter solutions in the appendix. We analyze the features of thermodynamic properties of the black hole solutions. The mass, Hawking temperature as well as thermal stability and the phase transitions are discussed. Moreover, the equation of state and critical phenomena associated with these solutions are also explored.
\end{abstract}
\maketitle
\section{INTRODUCTION}\label{1}
General Relativity (GR) \cite{adler1975introduction} is the simplest and most successful theory of gravitation. GR has been proposed for over one century and confirmed by many experiments \cite{will2014confrontation}. However, there are still
a few unknown issues such as dark energy and dark matter which indicate that GR may not be perfect yet. These issues strongly motivate us to consider the possibility of modified gravity theory beyond GR. Moreover, as demanded by string theory \cite{lust1989lectures}, black holes in higher dimensions must have higher curvature corrections. In the past decades, many possible schemes of modified gravity have been proposed \cite{clifton2012modified}. In most of these theories, one trouble is that we usually need to solve higher derivative equations. One way out is to consider a specific combination of the higher-order polynomial of the Riemann tensor such that the equation of motion remains two derivatives. The simplest choice of such kind of theory is the famous Gauss-Bonnet gravity \cite{lovelock1971einstein}. In addition, such quadratic Gauss-Bonnet term also appears as the one-loop correction of heterotic string theory \cite{gross1986superstring}.

The recent theoretical study proposes a scenario in which the fundamental building blocks of our Universe are extended objects rather than point objects \cite{1960Relativity}. The most natural and popular candidate is the one-dimensional strings object \cite{letelier1979clouds,herscovich2010black}. A cloud of strings behaves like a pressureless perfect fluid. Its one-dimensional character makes energy-momentum tensor only have a spacial component. Along this line, a lot of exact black hole solutions were obtained. Letelier \cite{letelier1979clouds} first presented a cloud of strings model extension of the Schwarzschild black hole. In this solution, the Schwarzschild radius of the black hole is thereby enlarged as $r_h=\frac{2M}{1-a}$ with $a$ being the string cloud parameter. This solution may have several astrophysical consequences \cite{richarte2008traversable}. Recently, many related papers considered the string cloud as a fluid in the spacetime background and constructed relevant exact solutions. For example, except in general relativity, the black hole solution with a cloud of strings in 4-dimensional Einstein-Gauss-Bonnet gravity was constructed in \cite{singh2020clouds}. Ghosh et.al \cite{ghosh2014cloud} discussed radiating black holes with a cloud of strings in second and third-order Lovelock gravity. Later, Rodrigues and Vieira constructed the Bardeen solution with a cloud of strings and studied its thermodynamic properties \cite{rodrigues2022bardeen}.

However, in most of these papers, the authors only consider black hole solutions without cosmological constant. Note that our current universe is undergoing an accelerating expansion which is the so-called dark energy issue. The simplest explanation for current cosmic acceleration is a nonzero positive cosmological constant \cite{peebles2003cosmological,Zhang2019a}. Moreover, the AdS/CFT correspondence usually requires us to consider the black hole solution with a negative cosmological constant. In addition, the AdS black holes usually possess much richer thermodynamical phase structure, for instance, the well known Hawking-Page phase transition can happen for the AdS black holes \cite{hawking1983thermodynamics}. Given the above motivations,  we will present a black hole in a cloud of strings background with a cosmological constant in Gauss-Bonnet gravity in higher dimensions. Both the uncharged and charged cases are considered and we will investigate the thermodynamic properties of these black holes in the presence of a cloud of strings. To enhance the readability of the paper, we only consider the positive Cosmological constant (dS) solutions in the main text and put the case of negative Cosmological constant (AdS) solutions in the appendix.

The structure of this paper is organized as follows: in SEC.\ref{2} we will solve the field equation and obtain (charged) black hole solutions coupled with a cloud of strings and cosmological constant; In SEC. \ref{3}, we will explore the thermodynamics properties including phase transitions and critical phenomena of the black hole with or without charge in dS space. Finally, we summarize our results in SEC. \ref{4}. In addition, in the appendix.\ref{5} we will briefly discuss the black hole solutions in AdS space. We will adopt units $8 \pi G=1$ and $\hbar = c = 1$.

\section{Gauss-Bonnet charged black hole solution}\label{2}
In this section we will derive Gauss-Bonnet charged black hole solution in dS spacetime surrounded by a cloud of strings (SC). The action of Gauss-Bonnet gravity with cosmological constant $\Lambda$ coupled to the Maxwell electrodynamics (ME) and a cloud of strings reads
	\begin{equation}
	S = \frac{1}{2}\int_{}d^4x\sqrt{-g}[R-2\Lambda+\alpha \mathcal{L}_{GB}+ \mathcal{L}_e(F^2)]+ S_{SC},
	\end{equation}
	where $g$ is the metric determinant, $R$ denotes the Einstein term and $L_e(F^2)$ represents the Maxwell electrodynamics Lagrangian with scalar $F = F_{\mu\nu}F^{\mu\nu}$, here $F_{\mu\nu}=A_{\mu;\nu}-A_{\nu;\mu}$. Moreover, $\alpha$ is Gauss-Bonnet coefficient and $L_{GB}$ denotes second-order Lovelock(Gauss-Bonnet) term, which is generally written as
	\begin{equation}
	\mathcal{L}_{GB} = R^2 - 4R_{\mu\nu}R^{\mu\nu} + R_{\mu\nu\lambda\sigma}R^{\mu\nu\lambda\sigma}
	\end{equation}
	and $S_{SC}$ is the action of a cloud of strings.
	The equations of motion can be obtained by $\frac{\delta S}{\delta g_{\mu\nu}} = 0$ and $\frac{\delta S}{\delta A_\mu} = 0$, in our case, that is
	\begin{equation}
	 	\label{einstain}
	 G_\nu^\mu + H_\nu^\mu - T_\nu^\mu =  0,
	\end{equation}
	\begin{equation}
	 	\label{conservation}
	 \nabla_\nu(F^{\mu\nu}\frac{dL_e(F^2)}{dF^2}) = 0,
	 \end{equation}
 	where $G_{\mu\nu}$ is Einstein tensor and $H_{\mu\nu}$ is tensor related to Gauss-Bonnet term, defined as,
 	\begin{equation}
 		H_\nu^\mu = -\frac{\alpha}{2}L_{GB}g_\nu^\mu + 2\alpha(RR^\mu_\nu-2R^{\mu\lambda}R_{\lambda\nu}-2R^\mu_{\lambda\nu\sigma}R^{\lambda\sigma}+R^{\mu\beta\lambda\sigma}R_{\nu\beta\lambda\sigma}).
 	\end{equation} $T_{\mu\nu}$ is the total energy-momentum tensor. Hence \eqref{einstain} reads
 	\begin{equation}
 		\label{total em}
 	G_\nu^\mu + H_\nu^\mu = -T_\nu^\mu(CS) + T_\nu^\mu(ME).
 	\end{equation}
 	Consider the following spherically symmetric metric ansatz in D dimension:
 	\begin{equation}
 		ds^2 = -f(r)dt^2 + \frac{1}{f(r)}dr^2 + r^2\gamma_{ij}dx^idx^j
 	\end{equation}
 	where $\gamma_{ij}$ is the metric of a $D-2$ dimensional constant curvature space $k = 1$. For the sake of convenience, we're going to use $n = D-2$ instead of dimension $D$, and we define
 	\begin{equation}
 		\Lambda = n(n-1)/2l^2
 	\end{equation}
 	The cosmological constant is positive, which indicates that the space is de Sitter. Now, \eqref{einstain} can be expressed as:
 	\begin{equation}
 	\frac{2r^2T^r_r}{n} = rf' + (n-1)(f^2 - 1) + (n+1)\frac{r^2}{ l^2} + 2\alpha(n-1)(n-2)\frac{1-f}{r}[f' - (n-3)\frac{1-f^2}{2r}] \\
 	\end{equation}
 	This differential equation is integrable, which can be written as:
 	\begin{equation}
 	[r^{n-1}(f-1)(1-\alpha(n-1)(n-2)\frac{f-1}{r^2})]' = -\frac{n+1}{ l^2}r^n + \frac{2}{n}r^nT^r_r
 	\end{equation}
 	Integrate the differential equation, we get the general metric function of Gauss-Bonnet black hole \cite{wiltshire1988black}:
 	\begin{equation}
 	f_{\pm} = 1 + \frac{r^2}{2\alpha(n-1)(n-2)}[1\pm\sqrt{1-4\alpha(n-1)(n-2)(-\frac{1}{ l^2}-\frac{\mu}{r^{n+1}}+\frac{2\mathcal{T}}{nr^{n+1}})}]
 	\end{equation}
 	Sign $\pm$ before the square root refers to the two different branches of metric solutions and \cite{boulware1985string} prove that the postive branch is unstable and the corresponding graviton has negative mass. Therefore, we only consider the negative branch; $\mu$ is a integration constant related to ADM mass via
 	\begin{equation}
 	\mu = \frac{32\pi M}{n(n-1)V_n},\quad V_n = \frac{2\pi^\frac{n+1}{2}}{\Gamma(\frac{n+1}{2})},
 	\end{equation}
	where $V_n$ is the volume of the n-dimensional unit sphere. $\mathcal{T}$ is associated with the total energy-momentum tensor by
 	\begin{equation}
 		\label{em}
 		\mathcal{T} = \int^{r}_{c}d\nu\nu^nT_r^r (\nu),
 	\end{equation} where an arbitrary constant $c$ leads to integration constant $\mu$. Next, we will discuss the energy-momentum tensor of the electromagnetic field and a cloud of strings respectively.

 \subsection{String cloud model}
 	Followed by the theory of a cloud of strings: the action of a cloud of strings is called Nambu-Goto action which is given by \cite{letelier1979clouds,singh2020clouds,rodrigues2022bardeen,gracca2018cloud}
	\begin{equation}
	S_{SC}=\int_{}\sqrt{-\gamma}md\lambda^0d\lambda^1,	
	\end{equation}
	where $m$ is a positive constant related to the string, $\lambda^0$ and $\lambda^1$ are timelike and spacelike parameters respectively. The string world sheet $\Sigma$ is determined by
	\begin{equation}
		\gamma_{AB} = g_{\mu\nu}(x)\frac{\partial x^\mu}{\partial \lambda^A}\frac{\partial x^\nu}{\partial \lambda^B},
	\end{equation}
	\begin{equation}	
		\gamma = det\gamma_{AB}.
	\end{equation}	
	Here we introduce the bivector associated with the string world sheet
	\begin{equation}
		\label{Sigma}
	\Sigma^{\mu\nu} = \epsilon^{AB}\frac{\partial x^\mu}{\partial \lambda^A}\frac{\partial x^\nu}{\partial \lambda^B},
	\end{equation}
	where $\epsilon^{AB}$ is the two-dimensional Levi-Civita symbol normalized as $\epsilon^{01} = -\epsilon^{10} = 1$. Within this setup, the Lagrangian density can be written as
	\begin{equation}
	\mathcal{L}_{SC}=m(-\frac{1}{2}\Sigma^{\mu\nu}\Sigma_{\mu\nu})^\frac{1}{2}.
	\end{equation}
	Therefore, the energy-momentum tensor for one string is
	\begin{equation}
	t_{SC}^{\mu\nu} = 2\frac{\partial L_{SC}}{\partial g_{\mu\nu}} = \frac{m\Sigma^{\mu\sigma}\Sigma_\sigma ^\nu}{(-\gamma)^\frac{1}{2}}.
	\end{equation}
	Replacing $m$ with proper density $\rho$, the energy-momentum tensor for a cloud of strings reads  \cite{letelier1979clouds,singh2020clouds,rodrigues2022bardeen,gracca2018cloud}
	\begin{equation}
	T_{SC}^{\mu\nu} = \frac{\rho\Sigma^{\mu\sigma}\Sigma_\sigma ^\nu}{(-\gamma)^\frac{1}{2}}.
	\end{equation}
	Quantities $\rho(-\gamma)^{1/2}$, $\rho\Sigma^{\mu\nu}$, $\Sigma^{\mu\nu}/(-\gamma)^{1/2}$ are manifestly gauge-invariant. The conditions for $\Sigma^{\mu\nu}$ to be a surface forming bivector are
	\begin{equation}
	\Sigma^{\mu\lbrack\alpha}\Sigma^{\beta\gamma\rbrack} = 0,
	\end{equation}
	\begin{equation}
	\nabla_\mu\Sigma^{\mu\lbrack\alpha}\Sigma^{\beta\gamma\rbrack} = 0,
	\end{equation}
	where the square bracket $[\alpha\beta\gamma]$ indicates anti-symmetrization of the indices. In conjunction with  difinition of the bivector \eqref{Sigma}, we get
	\begin{equation}
		\Sigma^{\mu\sigma}\Sigma_{\sigma\tau}\Sigma^{\tau\nu} = \gamma\Sigma^{\nu\mu}
	\end{equation}
	The combination the above identity with the conservation of energy-momentum tensor $T^{\mu\nu}_{;\nu} = 0$ leads to
	\begin{equation}
	\label{energy}
	\partial_\mu(\sqrt{-g}\rho\Sigma^{\mu\sigma}) = 0.
	\end{equation}
	Under the static spherically symmetric condition, the only nonzero component of the bivector $ \Sigma $ is $ \Sigma^{01} = -\Sigma^{10} $ and hence $ T^0_0(SC) = T^1_1(SC) = -\rho\Sigma^{01} $. Together with Eq. \eqref{energy}, we obtain $ \partial_r(\sqrt{r^{n}T^0_0}) = 0 $, which implies
	\begin{equation}
	T_\nu^\mu(SC) = \frac{a}{r^{n}}diag[1,1,0,...,0],
	\end{equation}
	where $a$ is string parameter, a positive integration constant that satisfies $(-\gamma)^{1/2}\rho = a/r^{n}$. In addition, $a$ should less than one so that the model has a clear meaning \cite{gracca2018cloud,singh2020clouds}.
	From \eqref{em} and \eqref{total em}, we have
	\begin{equation}
		\mathcal{T}_{SC} = -ar
	\end{equation}

 \subsection{Maxwell electromagnetic theory} 	
	As for electric term, the gauge field is $A_\mu = \phi(r)\delta_\mu^t$, under static condition \cite{mivskovic2011conserved,mivskovic2012thermodynamics}, where $\phi(r)$ denotes electric potential, and the electric field is $E(r) = -\phi'(r)$ with $'$ denotes the derivative respect to $r$. According to Maxwell electrodynamics, the field-strength is
	\begin{equation}
	F_{\mu\nu} = E(r)(\delta_\mu^t\delta_\nu^r-\delta_\nu^t\delta_\mu^r).
	\end{equation}
	Simple calculation shows $F^2 = -2E^2$, from \eqref{conservation}, we get
	\begin{equation}
	-\frac{d}{dr}(r^nE\frac{d\mathcal{L}}{dF^2}) = 0
	\end{equation}
	which results in the generalized Gauss law
	\begin{equation}
		\label{E}
	E\frac{d\mathcal{L}}{dF^2} = -\frac{q}{r^n}
	\end{equation}
	where $q$ is a parameter associated with charge by $Q^2 = 8V_n^2q^2$ \cite{pokhrel2023charged}. Based on Maxwell electric Lagrangian $\mathcal{L} = -F^2$, the Maxwell electric energy-momentum tensor takes the form
	\begin{equation}
	T_\nu^\mu(ME) = \delta^\mu_\nu\mathcal{L} - 4\frac{d\mathcal{L}}{dF^2}F^{\mu\lambda}F_{\nu\lambda}
	\end{equation}
	Now, from \eqref{E} we can obtain
	\begin{equation}
		E = \frac{q}{r^n} = \frac{Q}{\sqrt{8}V_nr^n}.
	\end{equation}
	The energy-momentum tensor then reads
 	\begin{equation}
 		T_\nu^\mu(ME) = -\frac{Q^2}{4V_n^2r^{2n}}diag[1,1,0,...,0]
 	\end{equation}
 	Therefore
 	\begin{equation}
 		\mathcal{T}_{ME} = \frac{Q^2}{4V_n^2(n-1)r^{1-n}}
 	\end{equation}
	So far, we are able to obatin the metric of Gauss-Bonnet charged black hole with cosmological constant in presence with a cloud of strings:
 	\begin{equation}
 		\label{f}
 	f = 1 + \frac{r^2}{2\alpha(n-1)(n-2)}[1-\sqrt{1-4\alpha(n-1)(n-2)(-\frac{1}{ l^2}-\frac{\mu}{r^{n+1}}-\frac{2a}{nr^n}+\frac{Q^2}{V_n^2n(n-1)r^2})}]
 	\end{equation}
 	In this paper, we shall stick to the common case $k = 1$ in dS space, thus the sign before $l^2$ is positive, and we will briefly discuss AdS space case in appendix.
	
\section{Thermodynamics}\label{3}
	In this section, we firstly discuss the thermodynamics of the dS black hole without charge \cite{cai2002gauss,cai2004gauss}. In this case, the metric function \eqref{f} reads
	\begin{equation}
		f = 1 + \frac{r^2 \left(1-\sqrt{1-4 \tilde{\alpha}  \left(-\frac{2 a r^{-n}}{n}-\frac{1}{l^2}-\mu  r^{-n-1}\right)}\right)}{2 \tilde{\alpha} }.
	\end{equation}  	
	Here $\tilde{\alpha} = \alpha(n-1)(n-2)$. Note that the event horizon locates at $f(r_+)=0$, we can express the ADM mass of the black hole in terms of the horizon radius $r_+$ as
	\begin{equation}
		\label{M}
	M(r_+) = \frac{(n-1) n V_n \left(r_+^{n-3} \left(\tilde{\alpha} -\frac{r_+^4}{l^2}+r_+^2\right)-\frac{2 a r_+}{n}\right)}{32 \pi }.
	\end{equation}
	Evidently, cloud of strings provides a source of negative mass, which may lead to the unstability of the system. Based on the black hole thermodynamics, the Hawking temperature $T$ is
	\begin{equation}
		\label{T}
	T = \frac{f'(r_+)}{4\pi} = \frac{r_+^{-2 n-1} \left(l^2 \left(n r_+^{2 n} \left(\tilde{\alpha} (n-3)+(n-1) r_+^2\right)-2 a  r_+^{n+4}-\right)-n (n+1) r_+^{2 n+4}\right)}{4 \pi  l^2 n \left(2 \tilde{\alpha}+r_+^2\right)}
	\end{equation}
	This expression is very complicated so that very hardly to do analytical study. In the following, we will use numerical analysis to explore the thermodynamic properties of dS black holes. First, we plot $T-r_+$ diagram:
	\begin{figure}[H]
		\includegraphics[width=0.49\textwidth]{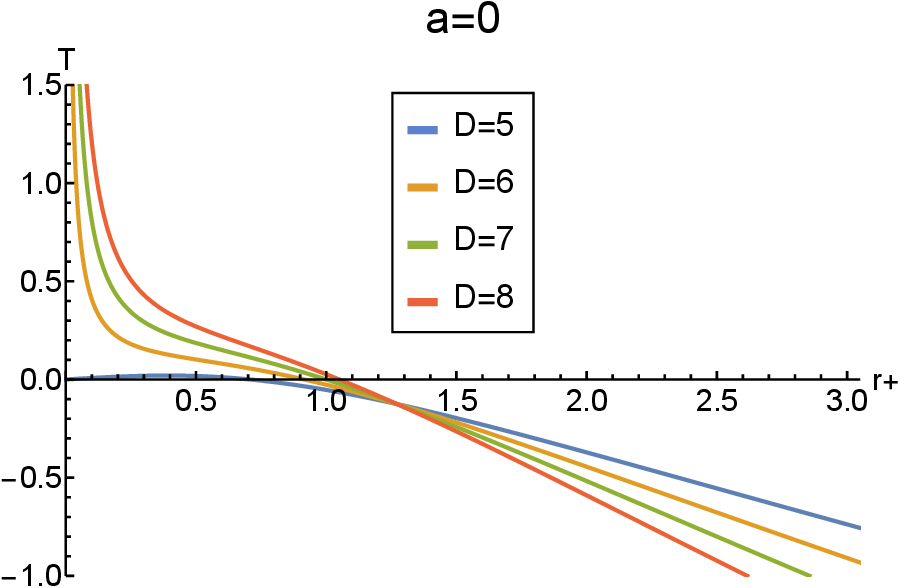}
		\includegraphics[width=0.49\textwidth]{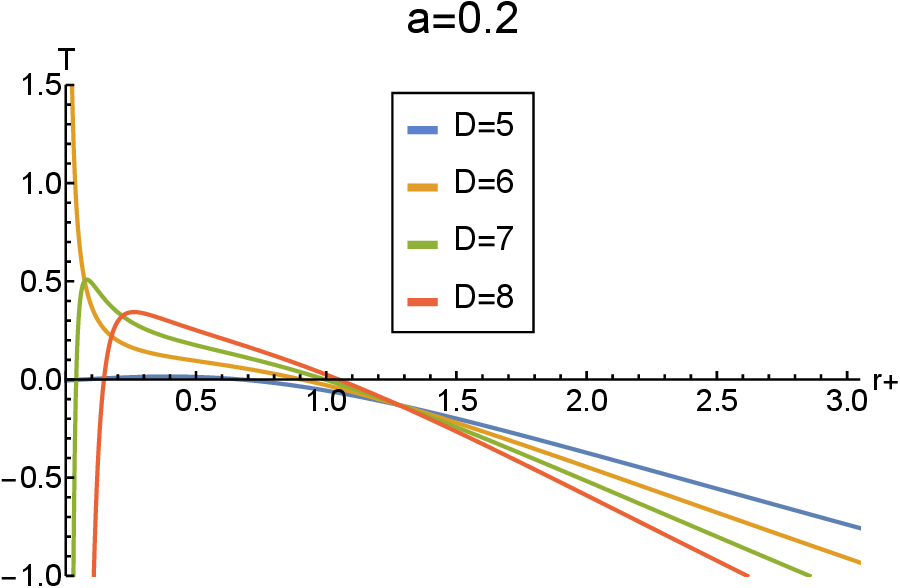}
		\caption{Graphic representation of Hawking temperature vs horizon radius with the values: $\alpha=1$, $l = 1$. Left: case in absence of cloud of strings, $a=0$; Right: case in presence of cloud of strings and $a=0.2$ }
		\label{Fig1}
	\end{figure}
	Note that negative temperature means unstable state. From the Fig. \ref{Fig1}, we can see that strings may cause the temperature to change from positive to negative in the extremely small radius region, in other words, cloud of strings make partially region unstable. Moreover, the first law of thermodynamics of the black hole $dM = TdS$ indicates that the entropy $S$ is
	\begin{equation}
	\label{entropy}
		S = \int{}{}dr\frac{1}{T}\frac{dM}{dr} = \frac{1}{8} (n-1) n V_n r_+^n \left(\frac{2 \tilde{\alpha}  }{(n-2) r_+^2}+\frac{1}{n}\right),
	\end{equation}
	which is independent of strings parameter $a$.

	\subsection{Thermodynamical stability}
	The thermodynamical stability includes local stability and global stability \cite{sancho1999global,chatzifotis2023thermal}, mainly determined by the behavior of heat capacity and free energy, respectively. Negative heat capacity means that the black hole is locally unstable. The divergence with a change of sign in heat capacity represents a second order phase transition, which the system changes from instability(stability) to stability(instability). The heat capacity of black hole is given by
	\begin{equation}
	C = \frac{\partial M}{\partial T} = \frac{\partial M}{\partial r_+}\frac{\partial r_+}{\partial T}.
	\end{equation}
	With the help of \eqref{M} and \eqref{T}, in our case, it reads
	\begin{equation}
			\label{C}
			C = \frac{C_1}{C_2},
	\end{equation}
	where
	\begin{equation}			
			C_1 = (n-1) n V_n \left(2 \tilde{\alpha}+r_+^2\right)^2\left(l^2 \left(2 a  r_+^{n+4} -n r_+^{2 n} \left(\tilde{\alpha} (n-3)+(n-1) r_+^2\right)\right)+n (n+1) r_+^{2 n+4}\right)  \nonumber
	\end{equation}
	and
	\begin{equation}
		\begin{aligned}
			C_2 =\ & 8 n r_+^{n+2} \left(2 \tilde{\alpha}^2 l^2 (n-3)+r_+^4 \left(6 \tilde{\alpha} (n+1)+l^2 (n-1)\right)+\tilde{\alpha} l^2 (n-7) r_+^2+(n+1) r_+^6\right) \nonumber \\
				  & -16 a  l^2 r_+^6 \left(2 \tilde{\alpha} (n-3)+(n-1) r_+^2\right).\nonumber
		\end{aligned}
	\end{equation}
	Then we plot $C-r_+$ diagram with different dimensions and parameter values:
	\begin{figure}[H]
		\includegraphics[width=0.5\textwidth]{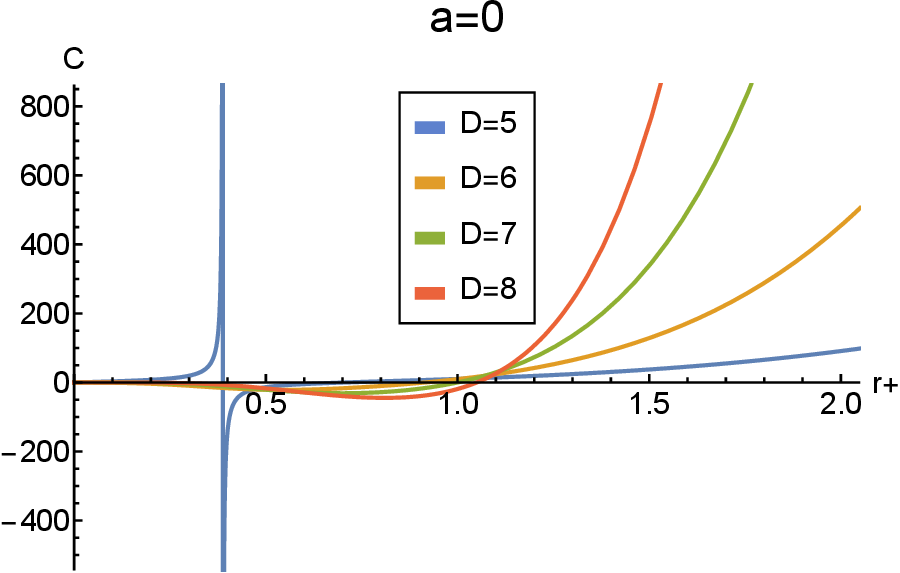}
		\includegraphics[width=0.5\textwidth]{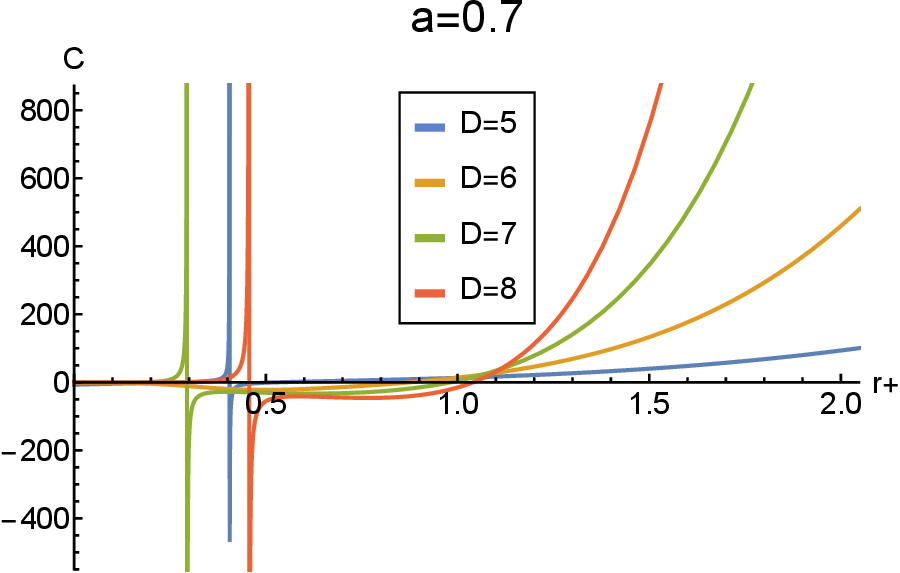}
		\caption{Graphic representation of heat capacity $C$ vs horizon radius $r_+$ with the values: $\alpha=1$, $l = 1$. Left: case in absence of a cloud of strings, $a=0$; Right: case in presence of a cloud of strings and $a=0.5$ }
		\label{Fig2}
	\end{figure}
	From Fig. \ref{Fig2}, there is originally no phase transition in seven and eight dimensions, but with the interference of the strings, new phase transitions occur. We can simply conclude that cloud of strings have a huge impact on the heat capacity: it increases the number of phase transitions. Moreover, we are interested in the global stability in dS space-time, positive free energy indicates that the black hole is globally unstable. The free energy is given by
	\begin{eqnarray}
		\label{G}
	F = M - TS = \frac{(n-1) V_n(F_1 + F_2)} {32 \pi }.
	\end{eqnarray}
	where
	\begin{equation}
	F_1 = -\frac{r_+^{-n-1} \left(\frac{2 \tilde{\alpha}}{(n-2) r_+^2}+\frac{1}{n}\right) \left(l^2 \left(n r_+^{2 n} \left(\tilde{\alpha} (n-3)+(n-1) r_+^2\right)-2 a  r_+^{n+4}\right)-n (n+1) r_+^{2 n+4}\right)}{l^2 \left(2 \tilde{\alpha}+r_+^2\right)}, \nonumber
	\end{equation}
	\begin{equation}
	F_2 = \tilde{\alpha} n r_+^{n-3}-\frac{n r_+^{n+1}}{l^2}+n r_+^{n-1}-2 a  r_+. \nonumber
	\end{equation}
	We plot the $F-r_+$ diagram as follows
	\begin{figure}[H]
		\includegraphics[width=0.5\textwidth]{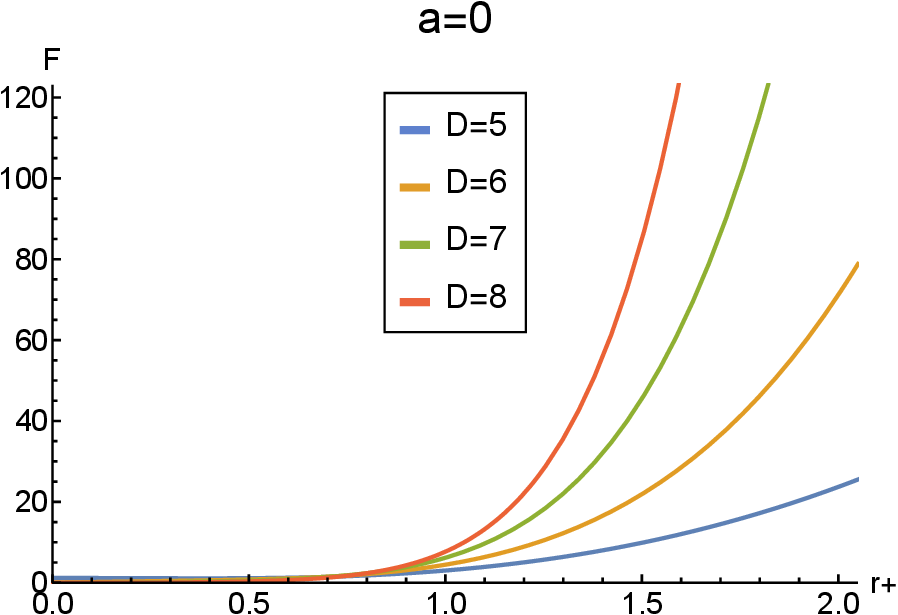}
		\includegraphics[width=0.5\textwidth]{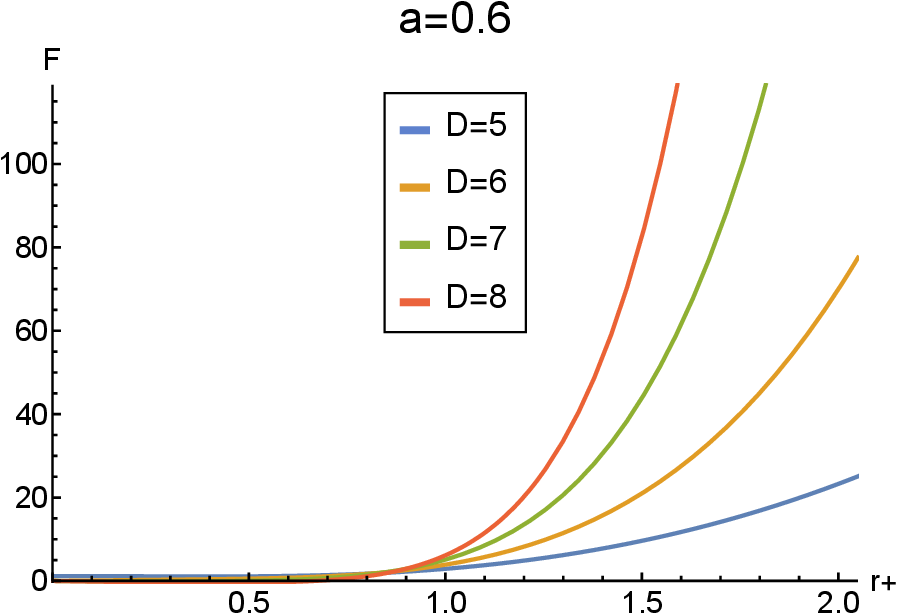}
		\caption{Graphic representation of free energy vs horizon radius with the values: $\alpha=1$, $l = 1$. Left: case in absence of cloud of strings, $a=0$; Right: case in presence of cloud of strings and $a=0.6$ }
		\label{Fig3}
	\end{figure}
	As can be seen from the Fig. \ref{Fig3}, the black hole is always globally unstable no matter with or without a cloud of strings.

	\subsection{Equation of state and critical phenomena}
	Thermodynamic pressure $P$ is thought to be associated with the cosmological constant by \cite{hendi2016extended,haroon2020thermodynamics}
	\begin{equation}
		\label{P}
		P= -\frac{\lambda}{8\pi} = \frac{n(n+1)}{16\pi l^2},
	\end{equation}
	and the volume reads
	\begin{equation}
		\label{V}
		V = \frac{\partial M}{\partial P} = \frac{(n-1) V_n r_+^{n+1}}{2 (n+1)}.
	\end{equation}
	It's exactly the thermodynamic volume of the system, thus it's different the geometric volume of the black hole. From \eqref{P}, \eqref{T}, we can write the equation of state as
	\begin{equation}
	P(T,r_+) = \frac{r_+^{-n-4} \left(n r_+^n \left(\tilde{\alpha} (n-8 \pi  r_+ T-3)+r^2 (n-4 \pi  r_+ T-1)\right)-2 a  r_+^4\right)}{16 \pi },
	\end{equation}
	The phase transitions happen in critical points, and the critical points occur at the inflection points of isothermal process that satisfies
	\begin{equation}
	\frac{\partial P}{\partial V} = 0,\qquad \frac{\partial^2 P}{\partial V^2} = 0,
	\end{equation}
	Or equivalently
	\begin{equation}
	\frac{\partial P}{\partial r_+} = 0,\qquad \frac{\partial^2 P}{\partial r_+^2} = 0.\label{critical}
	\end{equation}
	Solving Eq. \eqref{critical}, we can obtain the critical radius $r_c$ by equation
	\begin{equation}
	\frac{r_c^{-n-1} \left(-12 \tilde{\alpha}^2 (n-3) r_c^n+12 \tilde{\alpha} r_c^{n+2}+6 \alpha a  (n-3) r_c^4+(n-1) r^4 \left(a  r_c^2-r_c^n\right)\right)}{4 \pi  \left(6 \tilde{\alpha}+r_c^2\right) \left(12 \tilde{\alpha}+r_c^2\right)} = 0
	\end{equation}
	Take $n=3$ as a simple example, we can analytically solve the rational critical radius $r_c$ is
	\begin{equation}
		r_{c} = \frac{1}{2} \left(a\pm\sqrt{a^2+24 \tilde{\alpha} }\right)
	\end{equation}
	Notice that if Gauss-Bonnet coefficient is negative and string parameter $a$ is zero, the solution of $r_c$ is imaginary which means no critical phenomenon in the system. However, when strings parameter $a\neq 0$, the solution of $r_c$ could be real and the phase transition can happen. In other word, like the phase transition of heat capacity, cloud of strings may increase the number of phase transitions. Use critical radius $r_c$, we can get other critical values such as temperature $T_c$, pressure $P_c$ and volume $V_c$. Finaly, we succinctly discuss the thermodynamics properties of the black hole with charge. The thermodynamic quantities are given by
	
	(1). Entropy
	\begin{equation}
		S = \int{}{}dr\frac{1}{T}\frac{dM}{dr} = \frac{1}{8} (n-1) n V_n r_+^n \left(\frac{2 \tilde{\alpha}  }{(n-2) r_+^2}+\frac{1}{n}\right)
	\end{equation}
	It is the same as \eqref{entropy}, the entropy is also independent of charge.
	
	(2). Mass
	\begin{equation}
	M = \frac{(n-1) n V_n \left(-\frac{2 a r_+}{n}+r^{n-3} \left(\tilde{\alpha} -\frac{r_+^4}{l^2}+r_+^2\right)+\frac{Q^2 r_+^{1-n}}{2 (n-1) n V_n}\right)}{32 \pi }.
	\end{equation}
	Here, in contrast to strings parameter $a$, the charges $Q$ provides a source of positive mass.

	(3). Hawking temperature
	\begin{equation}
	T = -\frac{r_+^{-2 n-1} \left(l^2 \left(2 a r_+^{n+4}-n r_n^{2 n} \left(\tilde{\alpha}  (n-3)+(n-1) r_+^2\right)+\frac{Q^2 r_+^4}{2 V_n}\right)+n (n+1) r_+^{2 n+4}\right)}{4 \pi  l^2 n \left(2 \tilde{\alpha} +r_+^2\right)}.
	\end{equation}
	We plot $T-r_+$ diagram in the following:
	\begin{figure}[H]
		\begin{center}
		\includegraphics[width=0.6\textwidth]{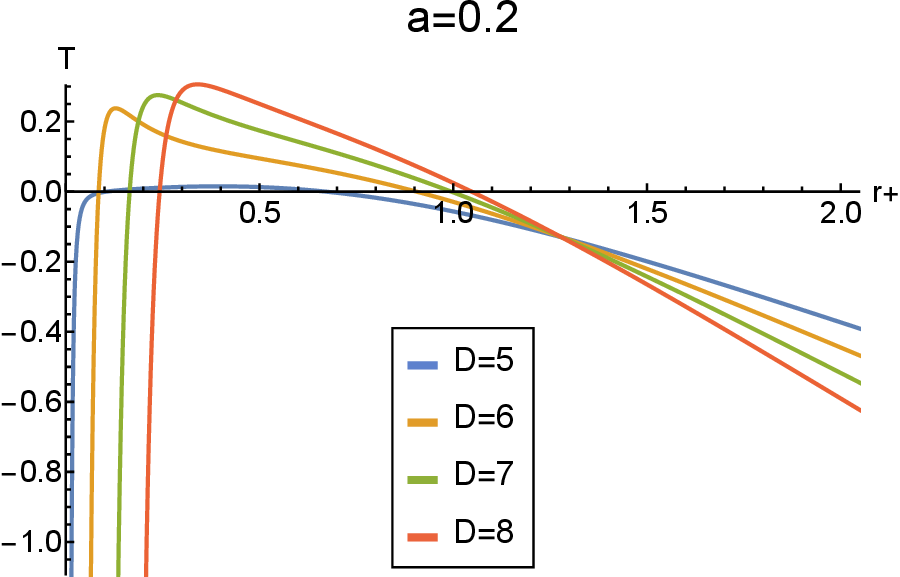}
		\caption{Plot of hawking temperature with respect of horizon radius with the values: $\alpha=1$, $l = 1$, $a=0.2$, $Q=0.01$.}
		\label{Fig4}
		\end{center}
	\end{figure}
	From the Fig. \ref{Fig4}, we can see that the charge behaves like strings, they also may cause the temperature to change from positive to negative in the small radius region.

	(4). Heat capacity: The form of the heat capacity is identical with \eqref{C}, but now
	\begin{eqnarray}
	C_1 &= &(n-1) n V_n \left(2 \tilde{\alpha} +r_+^2\right)^2 \nonumber\\
&&\left(l^2 \left(2 a r_+^{n+4}-n r_+^{2 n} \left(\tilde{\alpha}  (n-3)+(n-1) r_+^2\right)+\frac{Q^2 r_+^4}{2 V_n}\right)+n (n+1) r_+^{2 n+4}\right) \nonumber
	\end{eqnarray}
	\begin{equation}
		\begin{aligned}
			C_2 =\ &8 n r_+^{n+2} \left(l^2 \left(2 \tilde{\alpha} ^2 (n-3)+(n-1) r_+^4+\tilde{\alpha}  (n-7) r_+^2\right)+(n+1) r_+^4 \left(6 \tilde{\alpha} +r_+^2\right)\right) \nonumber \\
					&-16 a l^2 r_+^6 \left(2 \tilde{\alpha}  (n-3)+(n-1) r_+^2\right)-\frac{4 l^2 Q^2 r_+^{6-n} \left(-6 \tilde{\alpha} +4 \tilde{\alpha} n+(2 n-1) r_+^2\right)}{V_n} \nonumber
		\end{aligned}
	\end{equation}
	The $C-r_+$ diagram is as follows
	\begin{figure}[H]
		\begin{center}
		\includegraphics[width=0.6\textwidth]{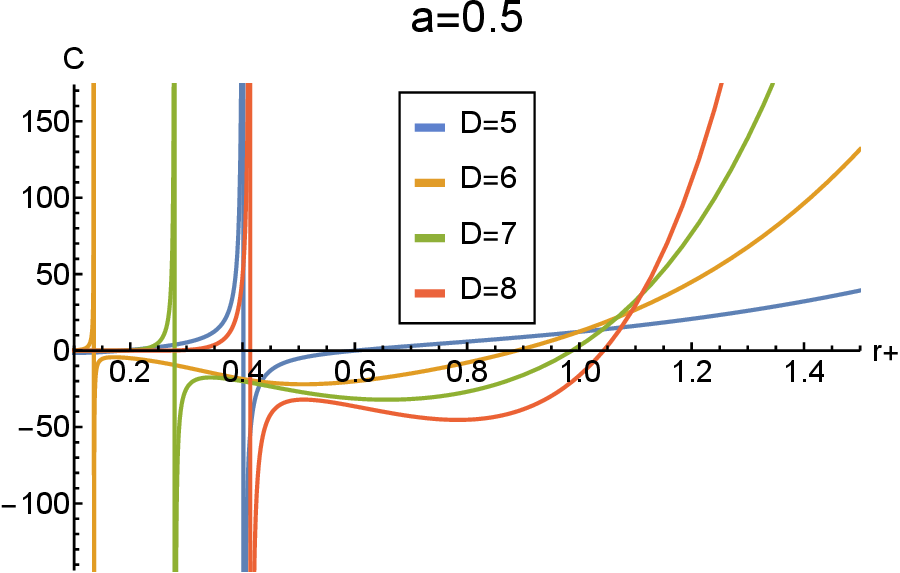}
		\caption{Plot of heat capacity with respect of horizon radius with the values: $\alpha=1$, $l = 1$, $a=0.5$, $Q=0.01$.}
		\label{Fig5}
		\end{center}
	\end{figure}
	Again, we can see that the effect of charges on heat capacity is similar to cloud of strings. With charges involved, now six dimension that haven't phase transition before have phase transition.

	(5). Free energy: The form of the free energy is identical with \eqref{G}, but with different $F_1$ and $F_2$ as
	\begin{equation}
		\begin{aligned}
			F_1 =\ & \frac{r_+^{-n-1}}{l^2 2 \tilde{\alpha} +r_+^2}  \left(\frac{2 \tilde{\alpha} }{(n-2) r_+^2}+\frac{1}{n}\right)[n (n+1) r_+^{2 n+4}+ \nonumber \\
				&l^2 \left(2 a r_+^{n+4}-n r_+^{2 n} \left(\tilde{\alpha}  (n-3)+(n-1) r_+^2\right)+\frac{Q^2 r_+^4}{2 V_n}\right)], \nonumber
		\end{aligned}
	\end{equation}
	\begin{equation}
	F_2 = -2 a r_++n r_+^{n-3} \left(\tilde{\alpha} -\frac{r_+^4}{l^2}+r_+^2\right)+\frac{Q^2 r_+^{1-n}}{2 (n-1) V_n}. \nonumber
	\end{equation}
	The $F-r_+$ diagram is as follows
	\begin{figure}[H]
		\begin{center}
		\includegraphics[width=0.6\textwidth]{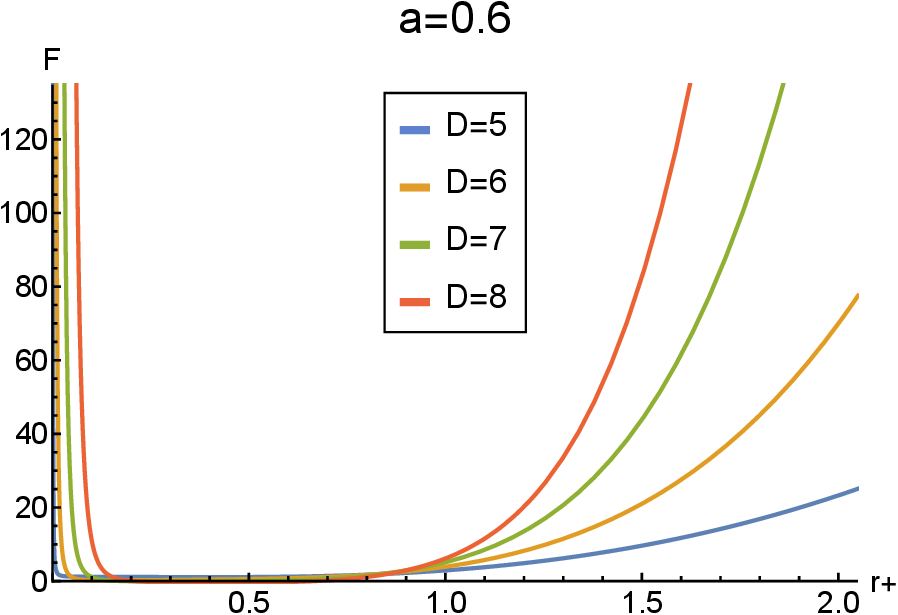}
		\caption{Plot of free energy with respect of horizon radius with the values: $\alpha=1$, $l = 1$, $a=0.6$, $Q=0.01$. }
		\label{Fig6}
		\end{center}
	\end{figure}
	Charges make the free energy of the small radius go to infinity, but it doesn't affect the sign of the free energy. Hence charges have only a numerical effect on the free energy. Charges also increase the area of negative temperature and the number of phase transitions, and the black hole is still globally unstable. The difference is that the charges provide a source of positive mass whereas strings provide a negative source for mass. The equation of state for the charged case reads
	\begin{equation}
		\begin{aligned}
			P(T,r_+) =\ &\frac{r_+^{-2 (n+2)}}{32 \pi  V_n } [-4 a  V_n r_+^{n+4}-Q^2 r_+^4  \\
			&-2 n V_n r_+^{2 n}\left(\tilde{\alpha} (-n+8 \pi  r_+ T+3)+r_+^2 (-n+4 \pi  r_+ T+1)\right)]
		\end{aligned}
	\end{equation}
	And the equation of critical radius $r_c$ is given by
	\begin{equation}
		\frac{f_1}{f_2}=0
	\end{equation}
	and
	\begin{equation}
		\begin{aligned}
			f_1 =\ &r_c^{-2 n-1} [ 2 V_n r_c^n \left(a (n-1) r_c^6+6 a \tilde{\alpha}  (n-3) r_c^4+r_c^n \left(-12 \tilde{\alpha} ^2 (n-3)-\left((n-1) r_c^4\right)+12 \tilde{\alpha} r_c^2\right)\right) \nonumber \\
			&+Q^2 r_c^4 \left(6 \tilde{\alpha}  (2 n-3)+(2 n-1) r_c^2\right) ] \nonumber
		\end{aligned}
	\end{equation}

	\begin{equation}
	f_2=8 \pi  V_n \left(6 \tilde{\alpha} +r_c^2\right) \left(12 \tilde{\alpha} +r_c^2\right) \nonumber
	\end{equation}

\section{CONCLUSIONS}\label{4}
	In this paper, we obtain the Gauss-Bonnet black hole solution with a cloud of strings in de-Sitter space. Both the uncharged and the charged cases are considered. By using the Einstein equation and static spherically ansatz, an exact black hole solution is obtained. When the string parameter $a=0$, our solution is reduced to the Gauss-Bonnet black hole solution in dS space \cite{cai2004gauss}. Then we focus on the effect of a cloud of strings on the thermodynamic properties of the black hole. We use the numerical method to calculate thermodynamic quantities and analyze them. We interestingly find that for the mass of the black hole, a cloud of strings provides a negative source for the black hole mass, while for temperature, a cloud of strings has only a quantitative influence.

	We also study thermodynamic stability based on heat capacity and free energy. We find that the existence of a cloud of strings has a significant impact on the stability of the system. The strings can increase the number of phase transitions due to heat capacity related to the strings. However, the globally unstable black hole is still retained even with the existence of strings.

	Moreover, we present the equation of state and discuss the critical phenomena. We express the conditions for the phase transitions and we can use these conditions to find critical points. We take the five-dimension as an example and discover that critical points correlate with strings, which indicates that the existence of phase transitions due to thermodynamic pressure is dependent on strings. The question of whether strings have any effect on other kinds of phase transitions will leave for future study.
\begin{acknowledgements}
This work is supported by NSFC with Grants No.12275087 and ``the Fundamental Research Funds for the Central Universities''.
	\end{acknowledgements}

\section{Appendix: Solutions in anti-de Sitter space}	\label{5}
	Here, we briefly discuss the Gauss-Bonnet black hole with a cloud of strings in AdS space: Now, the metric function $f$ reads
	\begin{equation}
		f = k + \frac{r^2}{2\alpha(n-1)(n-2)}[1-\sqrt{1-4\alpha(n-1)(n-2)\left(\frac{1}{l^2}-\frac{\mu}{r^{n+1}}-\frac{2a}{nr^n}+\frac{Q^2}{V_n^2n(n-1)r^2}\right)}]
	\end{equation}
	For AdS space, constant curvature space $k=1$, $-1$ or $0$ corresponding to the sphere, hyperbolic and flat geometry. For the sake of comparison with dS space, we only discuss $k=1$. Now, we can obtain all thermodynamic propeties. Here we plot Hawking temperature $T$, heat capacity $C$ and free energy $F$ with respect of horizon radius $r_+$ below:
	\begin{figure}[H]
		\includegraphics[width=0.5\textwidth]{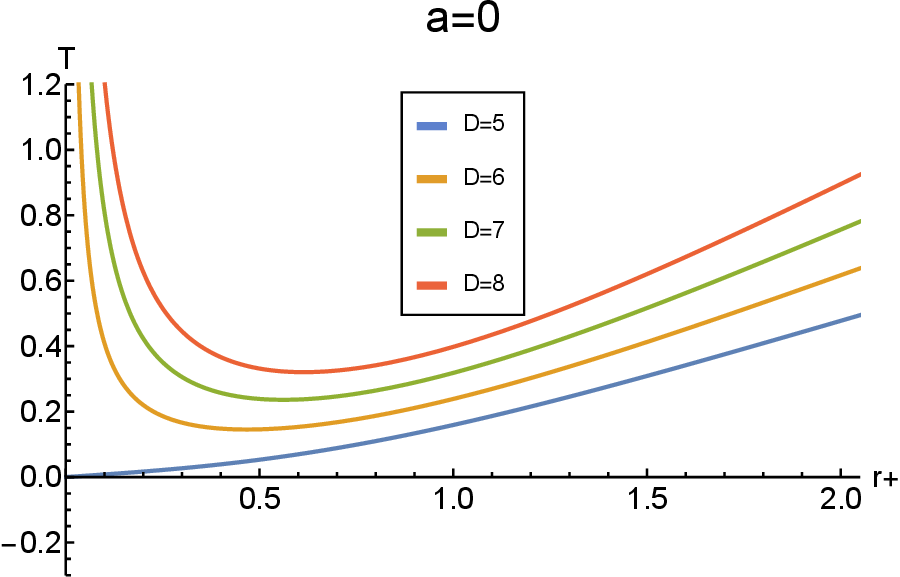}
		\includegraphics[width=0.5\textwidth]{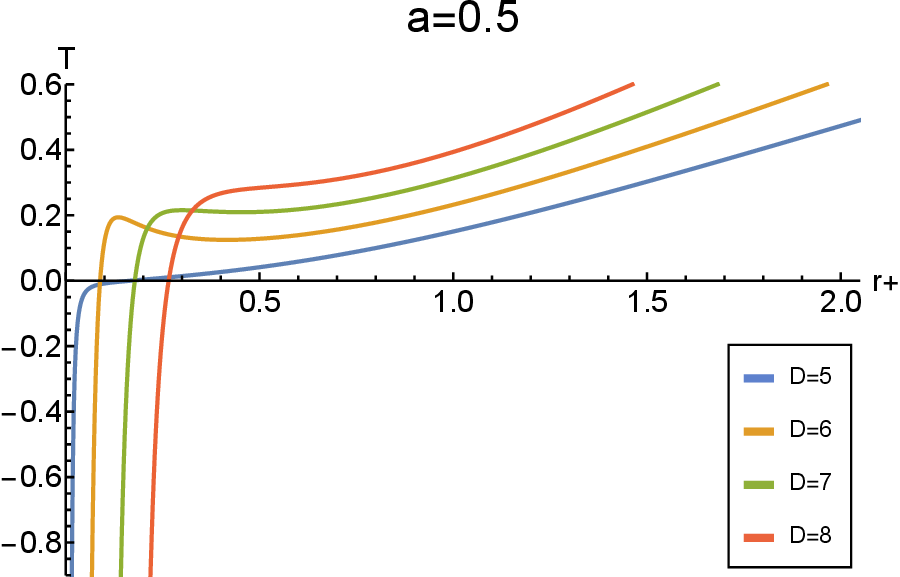}
		\includegraphics[width=0.5\textwidth]{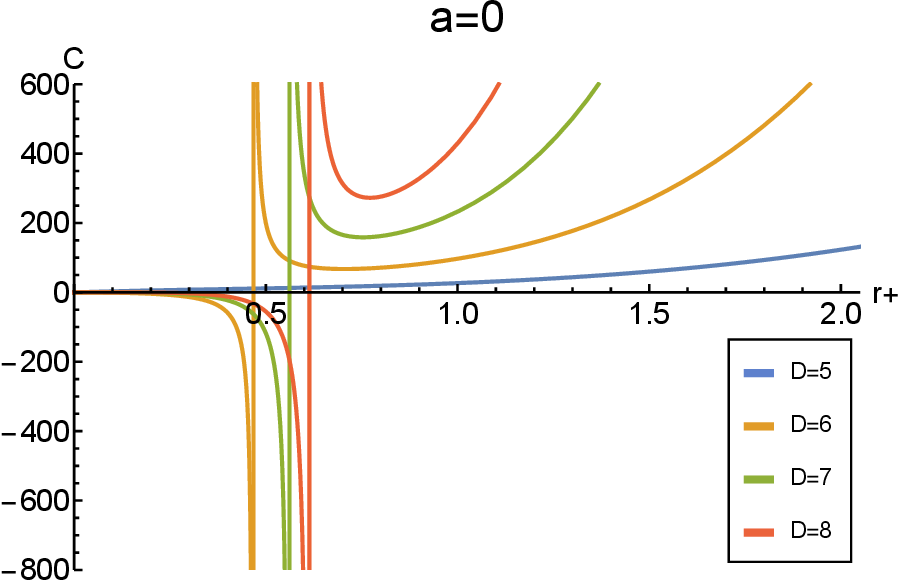}
		\includegraphics[width=0.5\textwidth]{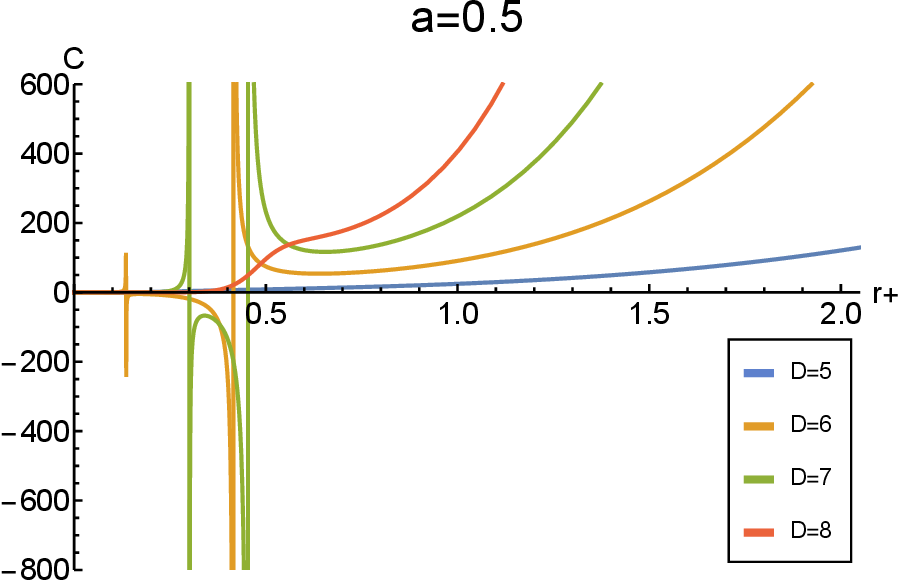}
		\includegraphics[width=0.5\textwidth]{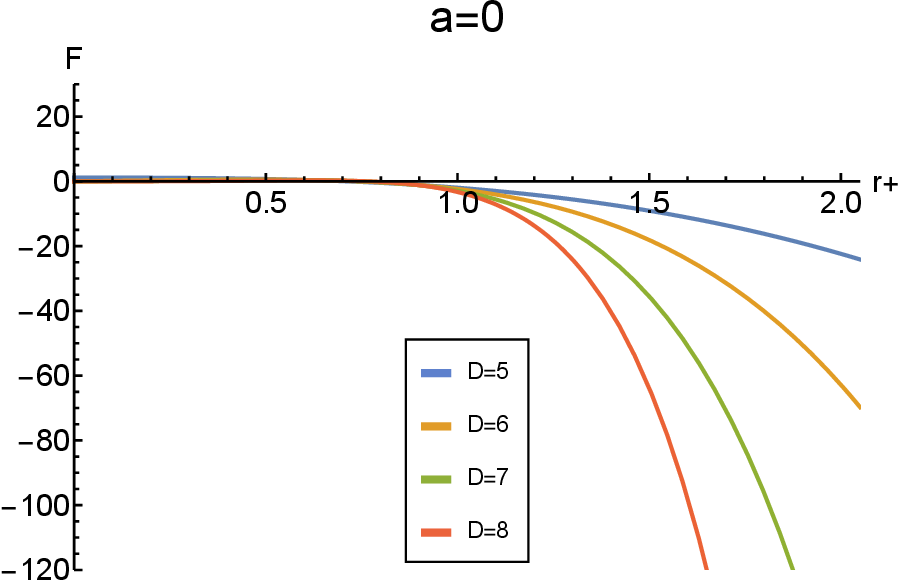}
		\includegraphics[width=0.5\textwidth]{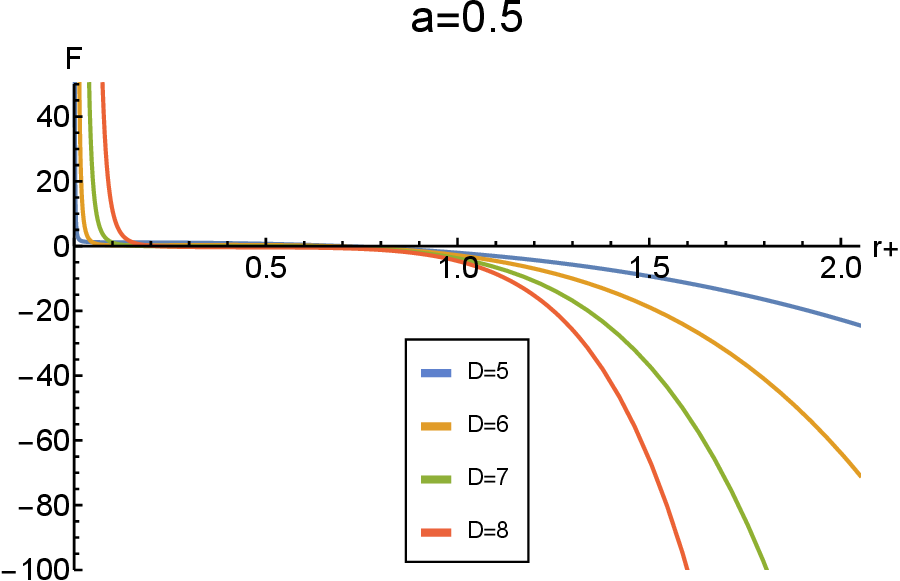}
		\caption{Plot of Hawking temperature, heat capacity and free energy (top to bottom respectively) vs horizon radius with the values: Left: $\alpha=1,l=1,a=0,Q=0$ ; Right:$\alpha=1,l=1,a=0.5,Q=0.01$ }
		\label{Fig7}
	\end{figure}
	Here we found that in AdS space, charge and cloud of strings not only change the property of the heat capacity, but also significantly affect the Hawking temperature. Charge and cloud of strings have an influence on the number of transitions like dS space. And temperature is constantly positive without strings and charges, but now the existence of charges and cloud of strings allows temperature to be negative. 

\bibliography{wenxian1}

\end{document}